\documentstyle[12pt]{article}
\twocolumn

\title{ATENSOR -- REDUCE program for tensor simplification
      }

\author{V.A.Ilyin\thanks{E-mail: ilyin@theory.npi.msu.su} {\small \it and} 
A.P.Kryukov\thanks{E-mail: kryukov@theory.npi.msu.su}\\ 
Institute of Nuclear Physics, Moscow State University\\
119899, Moscow, Russia}
\date{July 1996}

\begin{document}
\maketitle

\parindent -0mm

{\Large \bf PROGRAM SUMMARY}
\vskip 5mm

{\small
             {\it Title of program:} ATENSOR

             \vskip 1mm
             {\it Catalogue number:}

             \vskip 1mm
             {\it Program available from:} CPC Program Library, Queen's 
                 University of Belfast, N. Ireland (see application 
                 form in this issue).

             \vskip 1mm
             {\it License provisions:} none

             \vskip 1mm
             {\it Computer:} Any computers where REDUCE \cite{He} 
             can be installed

             \vskip 1mm
             {\it Operation system:} see above.

             \vskip 1mm
             {\it Programming language:} REDUCE-3.4, 3.5 

             \vskip 1mm
             {\it No. of bits in a word:} 32

             \vskip 1mm
             {\it No. of lines in distributed program, including test 
                  data, etc.:} 1963

             \vskip 1mm
             {\it Keywords:} Tensor, symmetry, multiterm linear identities,
             dummy indices, simplification, permutation group, REDUCE, 
                 computer algebra.

             \vskip 1mm
             {\it Nature of physical problem:} Simplification of tensor 
                 expressions with taking into account multiterm 
                 linear identities, symmetry relations and renaming dummy
                 indices. This problem is important for the calculations in the 
                 gravity theory,  differential geometry, other fields where 
                 indexed objects arise.

             \vskip 1mm
             {\it Method of solution:} The group algebra technique for 
                 permutation group is applied to construct a canonical subspace
                 and the effective algorithm for the corresponding projection.

             \vskip 1mm
             {\it Restriction on the complexity of the problem:} 
                 Computer operative memory is the severest restriction.

             \vskip 1mm
             {\it Running time:} It depends on the problem. For example the 
                 expression contained contraction of Riemann tensor with 
                 antisymmetric tensor of second order 
                 $\epsilon_{\mu \nu}*R_{\mu \nu \lambda \kappa} -
                  \epsilon_{\alpha \beta}*R_{\lambda \kappa \beta \alpha}$ 
                 require about 82s CPU time on HP9000/735 with 8M for REDUCE.

             \vskip 1mm
             {\it References:}

             [1] V.A.Ilyin and A.P.Kryukov, in Proc. of the Int. Symp. of 
                 Symbolic. and Algebraic
                 Computation (ISSAC'91), July 15-17, 1991, Bonn, Ed. by S.Watt, 
                 ACM Press (1991) 224.

             [2] V.A.Ilyin and A.P.Kryukov, in "New Computing techniques 
                 in Physics Research II",
                 Proc. of AIHENP-92, Ed. by D.Perret-Gallix, World Scientific,
                 Singapore (1992) 639-348
} 

\parindent 5mm

\vskip 1cm
\section*{\Large \bf LONG WRITE-UP}
\vskip 5mm

\section{Introduction}

    Objects with indices are often used  in  mathematics  and 
physics. Tensors are classical examples here \cite{LL,Tr}.  
Indexed  objects  can  have  very  complicated  and   intricated 
properties. For example the Riemann tensor has symmetry properties 
with respect to permutation of indices. Moreover it  satisfies 
the cyclic identity \cite{LL}. There  are a number of linear 
identities with many terms  in  the  case  of   Riemann-Cartan 
geometry with torsion \cite{Tr}. 

      So there is a problem of reduction of  express\-ions  which 
contain indexed objects, called "tensor expressions" below. 

     Two reduction strategies can be used. 

     First, the corresponding expressions are expanded in terms 
of basic elements to resolve symmetries and identities in  the 
explicit form. However, this way requires introducing a lot  of 
objects of different types and rules for their management.  In 
the Riemann tensor case, these are the Christofel symbol, the metric 
tensor, and their derivatives. As a rule, this  leads to  large 
intermediate expressions. Furthermore, such an approach 
fails, for instance, in the case of geometry with torsion.

     Second strategy is to consider indexed  objects  as 
formal objects with some properties. Note that if  we  consider  tensors 
which have only symmetries corresponding to  permutation  indices 
and renaming dummy ones then the problem can be solved  in  terms  of 
double cosets of permutation group \cite{RT}. However,  if  linear  
identities  with  many terms ($>2$) are present, this approach fails 
because the  summation operator leaves the group framework. 

Our approach to the problem of
simplification  of  tensor expressions is based on 
the consideration of tensor expressions as vectors in 
some linear space. The preliminary version of this idea was proposed in
\cite{I90}.
The advanced version of the algorithm was presented at ISSAC'91 \cite{IK91}
and AIHENP-92 \cite{IK92}. Here we present a program in which we implement 
the proposed algorithm in REDUCE and give a detailed description of the
program. We generalized the algorithm to the case of tensor multiplication.

Complementary approach was developed in \cite{Fulling}. Here Young 
diagram technique was used to solve the simplification problem in 
some specific case. Namely, when dummy indices are allowed only between 
basic tensors for which generic symmetry relations and multiterm 
linear identities should be imposed. 
Although this algorithm is
a powerful tool in `indicial tensor calculations' 
of asymptotic expansions of heat kernels of differential 
operators it fails in general case.

   From the user's point of view, there are three groups of tensor properties: 

\vskip 1mm

\noindent
{\bf S} - symmetry with respect to index permutations; 

\noindent
{\bf I} - linear identities. 

\noindent
{\bf D} - invariance with respect to renamings of dummy indices; 

\vskip 1mm

     As an illustration, for the Riemann curvature tensor these properties are: 

\vskip 1mm
{\small
\begin{tabular}{ll}
{\bf S:}	& $R_{abcd}=R_{cdab}, \qquad R_{abcd}=-R_{bacd};$ \\
{\bf I:}	& $R_{abcd}+R_{acdb}+R_{adbc}=0;$\\ 
{\bf D:}	& $R_{abcd} R_{ckmn} R_{dlps}=
 R_{abdc} R_{dkmn} R_{clps}.$
\end{tabular}
}
\vskip 1mm

Note that multiterm linear identities will produce many rewriting rules
which can complicate the problem essentially.
 
     The problem under investigation can be formulated as a question: 
{\it whether two tensor expressions are equal or not, taking into account 
S-I-D properties?} Then the problem of {\it simplest (shortest) canonical 
form for given expression arises as a central one.}

\section{Basic tensors and tensor expressions}   

Let us give some definitions.

Under {\it basic tensors} we will understand the object with finite number 
of indices which can have such properties as {\it symmetry} and 
{\it multiterm linear identities}\footnote{The symmetry 
relations are partial case of linear identities indeed}.

Then, under {\it tensor expression} we will understand any expression which can 
be obtained from basic tensors with the help of the following operations:
\begin{itemize}
\item {summation with integer coefficients;}
\item {multiplication (commutative) of basic 
tensors\footnote{Including contraction of indices.}.
      }
\end{itemize}

We assume that all terms in the tensor expression have the same number 
of indices. Some pairs of them are marked as dummy ones. The set of nondummy 
names have to be the same for each term in the tensor expression. The names
of dummies can be arbitrary.

\section{Algebraic approach}   

To start with,
  let us consider the case of one basic tensor and tensor expressions which
  are linear combinations of this basic tensor with integer coefficients
  (i.e. without multiplication of tensors).

     Let a tensor $F$ has indices $\mu_1, \ldots ,\mu_n$.
There are $ n! $ formally different  objects 
$$ F_{\pi(\mu_1, \ldots , \mu_n)},\quad 
 \pi = \left \{  
     \matrix{ 1& \cdots &n \cr \pi(1) & \cdots & \pi(n) \cr }
     \right \} \in S_n, $$

\noindent
where $ S_n $  is the group of permutations of the set ($ 1, \ldots ,n $) and 

$$\pi (\mu_1,\ldots,\mu_n)\equiv (\pi(\mu_1),\ldots ,\pi(\mu_n))$$.

     If $ F $  has symmetries with respect to index permutations it 
means that there is a subgroup $ H \in S_n $ such that 
$$ F_{h\circ\pi}\;-\;d(h)\cdot F_{\pi}\;=\;0,\qquad \forall h \in S_n ;$$
$$ d(h)\in R^1; \; h\circ\pi\equiv (\pi(h(\mu_1)), \ldots, 
\pi(h(\mu_n))). $$
         
     Multiterm linear identities can be written in the following form
$$ \sum_{\pi \in S_n} \ \alpha_{\pi} \cdot F_{\pi}=0, 
   \quad \alpha_{\pi} \in {\bf R}^{1}. $$

     If some pairs of dummy indices are present  without any loss of  
generality  we  may  suppose that their names are already normalized 
in some way, i.e. they have fixed names. Therefore one can only exchange
these names: 1) change names inside each pair and 2) change pairs of these 
names. With this restriction the exchanging of  dummy  indices  means 
that a subgroup $ M \in S_n $ exists such that\footnote{
The difference in actions of $ H $ and $ M $ on $ S_n $ in the
discussed relations ($ h \circ \pi $ and $ \pi \circ m$ ) is connected with 
the different nature of these transformations: the symmetry acts upon place of
indices while dummy indices renaming acts upon the names.}
$$  F_{\pi \circ m(\mu_1, \cdots , \mu_n)}= F_{\pi}, \quad
\forall\; m \in M, \pi \in S_n. $$

     Let us consider now {\it the group algebra} of $ S_n $ \cite{Na}. This
is a linear space $ {\bf R}^{n!} $ in which unit vectors correspond to 
permutations:  
$$ {\bf R}^{n!} \ni e_{\pi} \leftrightarrow \pi \in S_n. $$ 
 
    The vectors $e_{\pi}$ are orthogonal to each other in the Euclidean metric.

     So  we  have  an  explicit  isomorphism  between  tensor 
expressions and points in ${\bf R}^{n!}$: 
$$ \sum_{\pi} \alpha_{\pi} \cdot e_{\pi} \leftrightarrow
\sum_{\pi} \alpha_{\pi} \cdot F_{\pi}, $$

\subsection{Subspace K}

     In terms of ${\bf R}^{n!}$, the left hand sides of the {\bf S-I-D} 
relations correspond to the vectors 
\begin{equation}
e_{h \circ \pi} -d(h) e_{\pi}, \quad e_{\pi \circ m} -e_{\pi}, 
 \quad \sum_{k=1}^{n!} \alpha_{k} e_k,
 \label{eq:SIDvectors}
 \end{equation}

$$h\in H,\; m\in M,\;\pi\in S_n.$$

     These vectors span some subspace ${\bf K}\in {\bf R}^{n!}$.                                                     
We denote its dimension as $K$.

Now  we  can  split group algebra of $S_n$ 
into orthogonal components in terms of  the Euclidean  metric 
$$ {\bf R}^{n!} = {\bf K} \oplus {\bf Q}. $$

     It  is  obvious  that  all points  of $ {\bf R}^{n!} $ lying  in 
$ {\bf K} $ correspond to tensor expressions which are equal to  zero  due 
to the {\bf S-I-D} properties. Thus, any two points $ A $ and $ A' $ lying on
the plane parallel to $ {\bf K} $ correspond to equal expressions: 
$$  (A - A') \leftrightarrow  \sum_{\pi \in S_{n}} \alpha_{\pi}^{SID}
   \cdot F_{\pi} = 0. $$
            
     The $ (n!-K) $ dimensional  subspace  $ {\bf Q} $  could  be 
chosen as a set  of  canonical  elements: a point $ A_Q $ can be determined
as a canonical representative of the {\bf S-I-D} equivalence class of $ A $.
Then one can conclude that the problem of comparing
tensor expressions may be solved by comparing their canonical representatives.
Such an approach was developed in \cite{I90} where the Gramm-Schmidt 
orthogonalization  
procedure was used as a main technical method. However, this 
procedure requires  too  much 
time and computer  memory  during  executions.  

    In \cite{IK91} the authors proposed an effective procedure where another
subspace, denoted below as {\bf L}, had been considered as a set of canonical 
elements. In the next section we give a concise formulation of this 
procedure.

\subsection{"Triangle" {\bf S-I-D} basis}

     Let's designate vectors (\ref{eq:SIDvectors}) as 
\begin{equation}
 V_k^0= \sum_{j=1}^{n!} \alpha_{kj}^0 e_j, \; k=(1,...,\tilde K).   
\label{eq:SIDvecV}
\end{equation}

  Then we construct new vectors by recurrent applying (with steps 
$m=1,\ldots ,\tilde K$) of the following transformation:

--- {\it if $V^{m-1}_m \neq 0$ then define $k_m$ by first nonzero coefficient
in $V^{m-1}_m$ and let}
$$ V_k^{m} \equiv V_k^{m-1},\qquad k=1,\ldots,k_m; $$
\begin{equation}
 V_k^{m} \equiv V_k^{m-1} - \sum_{i=1}^m 
  {\alpha_{kj_{i}}^{m-1} \over \alpha_{k_ij_i}^{m-1}} 
  V_{k_i}^{m-1}, \; k>k_m;
							\label{eq:triangK}
\end{equation}
--- {\it if $V^{m-1}_m =0$ then $V_k^{m} \equiv V_k^{m-1}$.}

     As a result, we have a set  of nonzero vectors  $ V_k^{(K)} $
which span the subspace $ {\bf K} $. Note that all nonzero vectors 
$ V_k^{(K)} $ are linearly independent. Indeed, let us reorder the unit vectors
$$ (e_1, \ldots ,e_{n!}) \longrightarrow (e_{j_1}, \ldots ,e_{j_m}, \ldots ). $$
Then the vectors $ V_k^{(K)} $ with $k\geq k_{m} $ will have zero 
projections onto the unit  vectors  $ e_{j_i}$ ($i<m$).  
In other words, the set $ V_k^{(K)} $ has a "triangle" form 
in this reordered basis. It is evident that a number of nonzero vectors
$ V_k^{(K)} $ is equal to $K$ - the dimension of the {\bf S-I-D} subspace
{\bf K}.

    We have made some optimization in addition to the transformation 
(\ref{eq:triangK}):

{\it on each step $m$ all already constructed nonzero vectors 
$V_{k_{i}}^{(K)}$ ($i<m$) are improved by the following transformation}:
 
\begin{equation}
 V_{k_i}^{m} \rightarrow V_{k_i}^{m} - { \alpha_{k_i j_m }^{m} \over 
      \alpha_{k_m  j_m}^{m} } V_{k_{m}}^{m}, \quad i<m.
							 \label{eq:Kopt}
\end{equation}

     Such optimization doesn't spoil the "triangle"  structure and vectors 
$ V_i^{k} $ with $ k<k_m $  will have zero  projections  onto  the 
unit vectors $ e_j $  with $ j=j_1,...,j_{m-1} $.
Thus, we conclude that each vector $ V_k^{(K)}$ 
has not more then $ (n!-K) $ nonzero components.

\subsection{Subspace of canonical elements}
  Let us consider an arbitrary vector, designated below as
$$ A^{0} \equiv A= \sum_{j=1}^{n!} a_i^{0} \cdot e_j, $$

\noindent
and recurrently apply ($m=1,\ldots,K$) the following transformation: 

\begin{equation}
A^{m} \equiv A^{m-1}- {a_{j_m}^{m-1} \over \alpha_{k_mj_m}^{m-1} }
  V_{k_m}^{(K)} = \sum_{j=1}^{n!} a_{j}^{m} e_j.
						\label{eq:CanProj}
\end{equation}

It is easy to see that this linear transformation has the following properties:

\begin{itemize}
\item[--]  it shifts the vector  $A$ 
along the plane parallel to $ {\bf K} $. ({\it So $A$ and $A^{K+1}$
correspond to equal tensor expressions}); 
\item[--]  vector $A^{K+1} $ has zero projections onto unit vectors $ e_j $  
with $ j=j_1, \ldots ,j_K $ ;
\item[--]     $A^{K+1}=0  \ \ \ $ iff $ \ \ \  A \in {\bf K}. $
\end{itemize}

Thus, we  conclude that $\forall A \in {\bf R}^{n!}$:

\begin{itemize}
\item $A^{K+1} = {A'}^{K+1} $ iff $ (A-A') \in {\bf K}; $
\item transformation (\ref{eq:CanProj}) doesn't shift $A^{K+1};$ 
\item image of the transformation (\ref{eq:CanProj}) 
 is some linear subspace $ {\bf L} $ with dimension $ (n!-K) $ and 
$ {\bf L} \cap {\bf K}= \{0\}$\footnote{Here  $\{0\}$ 
means that this set has a single point - zero vector.}; 
\item every vector in $ {\bf L} $ has not more than $ (n!-K) $ nonzero components.
\end{itemize}

{\bf  We define the subspace $ {\bf L} $ 
 as a set of cano\-ni\-cal re\-pre\-sen\-ta\-tives for {\bf S-I-D} equivalence
classes under consideration}.

\section{Multiplication}			\label{Mult}

If a tensor expression is obtained by the multiplication of basic tensors
then  we directly generate  the set of {\bf S-I-D} relations
as a {\it product} of basic ones.

 Then the expression is elaborated,
taking into account additional relations originated from the multiplication
rule (in our case from the commutativity). 
Let us consider a tensor $tt$ with two indices as an example. If we
multiply it by itself, say as $tt(i,j)*tt(k,l)$,
the additional symmetry appears: 
$$tt(i,j)*tt(k,l) = tt(k,l)*tt(i,j).$$

We consider all such permutations and add the corresponding
elements to the {\bf S-I-D} relations. Then we perform the "triangle"
procedure (\ref{eq:triangK},\ref{eq:Kopt}) to construct the full {\bf K}
subspace, and use the procedure (\ref{eq:CanProj}) for the canonical 
representative calculation\footnote{It is clear how to generalize the algorithm
to the case of a noncommutative finite algebra of
basic tensors. In such cases, more complicated relations will appear
instead of the simple symmetry written above. However, in any case they will
be linear identities which can be elaborated naturally in the 
framework of our approach.}.

\section{Some definitions}

 Let us give some definitions  which we use in the following.
\begin{enumerate} 

\item ${\bf K}$-{\it basis} is a general name for "triangle" set of linear 
independent vectors $V^{({\bf K})}_k$ ($k=k_1,\ldots ,k_K)$.

\item ${\bf K_0}$-{\it basis} is the ${\bf K}$-basis for a
  basic tensor considered as a separate tensor expression, taking into account
 its symmetries and linear identities only.

\item ${\bf K_M}$-{\it basis} is the ${\bf K}$-{\it basis} of the expression 
under consideration which arises
  from ${\bf K_0}$-{\it bases} of basic tensors and 
  relations generated by their multiplication.

\item ${\bf K_D}$-{\it basis} is the completion of the ${\bf K_M}$-{\it basis}
up to the full ${\bf K}$-{\it basis} of the expression under consideration, 
taking into account relations which arise from renamings of 
  dummy indices. We shall also call this basis the
  {\it full} ${\bf K}$-{\it basis}.

\item To {\it sieve} some ${\bf S-I-D}$ vector means to do the step $m$ 
of the "triangle" procedure (\ref{eq:triangK}).

\item To {\it rearrange} some ${\bf S-I-D}$ vector means to do the step $m$ 
of the "triangle" procedure (\ref{eq:Kopt}).

\item To {\it sieve} some vector $A$ by a ${\bf K}$-{\it basis}
means to apply the transformation (\ref{eq:CanProj}).
\end{enumerate}

\section{Algorithm}

There are two sets of operations. One is performed due to the introduction
of new basic tensors, and the result of these operations is the construction
of the new ${\bf K_0}$-bases. Another set of operations is connected with the 
simplification of tensor expressions itself.

The starting procedure for the construction of ${\bf K_0}$-bases
is $TSYM$ (section \ref{tsym}), and the algorithm is:

\begin{enumerate}
\item {Generate full list of {\bf S-I} vectors (\ref{eq:SIDvectors}): 
        $L_{\bf S-I}=\{v\}$.}
\item {Let ${\bf K} = {\bf K_0}$ where ${\bf K_0}$ is initial basis
        (may be empty).}
\item {If the list $L_{\bf S-I}$ is empty then finish.}
\item {Take the next vector $v$ from $L_{\bf S-I}$ and delete it from
           this list: $L_{\bf S-I} \rightarrow L_{\bf S-I}/\{v\}$.}
\item {Sieve the vector $v$ through the ${\bf K}$-basis (transformation 
    (\ref{eq:triangK}), procedure $sieve\_pv$, section \ref{WorkWithK}).} 
    The result is a new vector $v'$.
\item {If $v' \ne 0$ then insert $v'$ in the
      ${\bf K}$-basis and rearrange the basis (transformation (\ref{eq:Kopt}),
          procedure $insert\_pv$, section \ref{WorkWithK}).}
\item {Repeat from step 3.}
\end{enumerate}

All ${\bf K_0}$-bases are stored. There is a possibility to delete any 
basic tensor
from the list of tensors and so to delete its ${\bf K_0}$-basis (section
\ref{TClear}).

The algorithm of simplification of tensor expressions is:
\begin{enumerate}
\item If there is no multiplication of tensors or there are no new 
  {\bf S-I} relations due to their multiplication then go 
  to step `{\it 4}'.
\item Expand the ${\bf K_0}$-bases involved up to the necessary rank of
  the permutation group corresponding to the multiplication of basic
  tensors in the expression elaborated (procedure $t\_upright$, 
  section \ref{TMult}), and collect the obtained relations as the initial
  ${\bf K}$-basis of the considered expression. 
\item Complete the ${\bf K}$-basis by additional vectors 
      which arise from the multiplication (see section \ref{Mult}) 
      by the algorithm of the ${\bf K}$-basis construction (see above).
  The result is the ${\bf K_M}$-basis for the expression 
  under simplification.
\item {\it Dummies.} Complete the ${\bf K_M}$-basis by vectors corresponding
   to the relations which arise from renamings of dummy indices
   (if they are present).
   The result is the ${\bf K_D}$-basis - the full ${\bf K}$-basis
   for the expression under simplification. 
\item Sieve the tensor expression through the ${\bf K_D}$-basis 
   (transformation (\ref{eq:CanProj}), 
   procedure $sieve\_t$, section \ref{WorkWithK}). The result is the
   construction of the canonical representative for the expression.
\end{enumerate}

Any ${\bf K_M}$ and ${\bf K_D}$-bases are not stored and are constructed
each time an expression is simplified.

\section{Program description}

ATENSOR program consists of the following blocks:

\begin{itemize}
\item  {interface with REDUCE system;}

\item  {generator of permutations;}

\item  {p-vector arithmetics;}

\item  {tensor arithmetics;}

\item  {generator of the {\it multiplication} relations;}

\item  {generator of the {\it dummy} relations;}

\item  utilities to work with $\bf K$-bases.
\end{itemize}

\subsection{Interface with REDUCE}

Interface with REDUCE is implemented by defining of a domain \cite{BHP}. 
This is a natural way to implement a new object in REDUCE. We should define 
the following set of procedures:

\vskip 0.5cm
{\small
\noindent
\begin{tabular}{|l|l|l|}
\hline
Operation & Internal proc. & Comment \\
\hline
minus	&	t\_minus&	Unary minus\\
plus	&	t\_plus	&	Summ\\
times	&	t\_times&	 Product\\
difference&	t\_difference&	Substruction\\
zerop	&	t\_zerop&	Does tensor \\
        &               &       equal zero?\\
\hline
prepfn	&	t\_prep&\\
prifn	&	t\_pri	&	Print function\\
intequiv&	t\-intequiv&	Is tensor equi- \\
        &                  &     valent to \\
         &                &   integer\\
\hline
\end{tabular}
}
\vskip 0.5cm

The following procedures must be define for completeness but can not 
be used as operations for tensors. 
These procedures produce an error message if called.

\vskip 0.5cm
\noindent
{\small
\begin{tabular}{|l|l|l|}
\hline
Operation & Internal proc. & Comment \\
\hline

expt	&	t\_expt	&	Power\\
quotient&	t\_quotient&\\
divide	&	t\_divide&\\
gcd	&	t\_gcd	&	Great common \\
      &                 &     divider\\
minusp	&	t\_minusp&	Is tensor \\
        &             &       negative?\\
onep	&	t\_onep	&	Does tensor \\
    &              &equal 1?\\
i2d	&	i2tensor&	Transform \\
      &                &integer \\
       &                & to tensor \\
\hline
\end{tabular}
\vskip 0.5cm

We also define the domain name ($TENSOR$) and the tag ($!\!:\!TENSOR$). 

To complete the interface with REDUCE, we add
the tag of the tensor domain to the global variable $DOMAINLIST!\!*$

This method supports the input process of tensor expressions and provides 
calls of the corresponding internal procedures automatically.

\subsection{Permutations}

This block implements generation of permutations of $N$ order and includes 
some procedures for working with them. All procedures work with the
{\it packed} and the {\it unpacked} form of permutations.

{\it Unpacked} form of a permutation p is a list of numbers\footnote{see 
section \ref{UseMem} for details.}: 
$$p=(d_1\; d_2\; ...\; d_k),\quad  1 \leq d_i \leq k \leq 99. $$  

{\it Packed} form is the corresponding number: 
$$p=d_1d_2...d_k. $$

For example,
$$p=(1\; 2\; 3\; 4\; 5) \qquad  \mbox{and} \qquad  p=12345. $$

The transformation of a permutation from one form to another is made
automatically. The packed form is more economic with respect to computer 
memory, but requires more time to proceed. There is the global variable 
$!\!*\!ppacked$. If it is $T$ (the default value) then all permutations are
packed; if $NIL$ then they are stored in the unpacked form.

We use the well known algorithm for generation of permutations 
\cite{Kn}. We implement the following procedures:

\begin{itemize}
\item  {$mkunitp(n)$ -- generates the unit element of $S_n$};
\item  {$pfind(p_1,p_2)$ -- returns $x$ such that $p_2 = x\circ p_1$};
\item  {$prev(p)$ -- returns reverse permutation $x$ such that $x\circ p=1$};
\item  {$psign(p)$ -- returns $(-1)^k$, where $k$ is the number of 
         transpositions which are necessary
         to apply to the permutation p to get the identical permutation.
       }
\item  {$pmult(p_1,p_2)$ -- returns the permutation $x=p_1\circ p_2$};
\item  {$pappl(p,l)$ -- returns $l$ with the elements permuted by $p$,
         so that $l$ is replaced by $p(l)$.} 
\end{itemize}

There are some utilities to work with permutations:

\begin{itemize}
\item[--]  {$pupright(p,d)$ -- extends the permutation $p\in S_n$
   to the right up to the element of $S_{n+d}$ with the identical permutation 
   of the extra indices
   (this utility is used for the elaboration of multiplication 
   of basic tensors);}  
\item[--]  {$pupleft(p,d)$ -- extends the permutation $p\in S_n$
  to the left up to the element of $S_{n+d}$ with the identical permutation 
  of the extra indices 
   (this utility is used for the elaboration of multiplication 
   of basic tensors);}  
\item[--]  {$pappend(p_1,p_2)$ - concatenates the permutation $p_1\in S_{n_1}$ 
  with the permutation $p_2\in S_{n_2}$. Returns the element of $S_{n_1+n_2}$; }
\item[--]  {$pkp(p)$ -- packs the permutation $p$;}
\item[--]  {$unpkp(p)$ -- unpacks the permutation $p$.}
\end{itemize}

\subsection{P-vectors}

$P$-vectors are one of the main objects in the program. 
They represent the vector in ${\bf R^{n\!}}$ --
the group algebra of $S_n$. In the program they are implemented as a
REDUCE domain \cite{BHP}. 

Internal structure of $p$-vector is: 

\vskip 1mm
{\small
\noindent
$p\!-\!vector$	::= $ (!\!:\!pv\; .\; p\!-\!list)$ \\ 
$p\!-\!list$	::= $ NIL|(coeff\;.\;perm)\;.\;p\!-\!list$ \\
$coeff$ ::=  $ integer$
}
\vskip 1mm

All the standard operations are defined for $p$-vectors because they form
a domain.

The following procedures are used for tensor simplification:

\begin{itemize}
\item  {$pv\_sort(pv)$ -- sorts the $p$-list so that all permutation 
       will be ordered, e.g. $p_i>p_j\; \forall\; i<j$;
       }
\item  {$pv\_compress(pv)$ -- removes all terms with zero coefficient;}
\item  {$pv\_renorm(pv)$ -- reduces the first coefficient (in integer numbers),
       i.e. \\$pv \rightarrow pv/GCD(c_1,c_2,...)$ 
        where $c_i$ are the coefficients.
       }
\end{itemize}

\noindent
Some utilities are available to work with $p$-vectors:

\begin{itemize}
\item[--]  {$pappl\_pv(p,pv)$ -- applies the permutation $p$
           to the $p$-vector $pv$. \\
           Returns $p' = \sum_i c_i\; pmult(p,p_i)$ 
           where $pv = \sum_i c_i\; p_i$;}
\item[--]  {$pv\_applp(pv,v)$ -- applies the $p$-vector $pv$ to the 
           permutation $p$. \\
           Returns $p' = \sum_i c_i\; pmult(p_i,p)$ 
           where $pv = \sum_i c_i\; p_i$;}
\item[--]  {$pv\_upright(pv,d)$ - expands the $p$-vector $pv$ to the right. \\
           Returns $p' = \sum_i c_i\; pupright(p_i,d)$ 
           where $pv = \sum_i c_i\; p_i$;}
\item[--]  {$pv\_upleft(pv,d)$ - extends the $p$-vector $pv$ to the left. \\
           Returns $p' = \sum_i c_i\; pupleft(p_i,d)$ 
           where $pv = \sum_i c_i\; p_i$.}
\end{itemize}

\subsection{Tensors}

Tensors are the main objects in the program. They represent tensor expressions. 
In the program they are implemented as a REDUCE domain \cite{BHP}. 

The internal structure of a tensor is: 

\vskip 2mm
{\small
\noindent
$tensor$ ::=  $ (!\!:\!tensor\; i\!-\!tensor_1\; ...\;\; i\!-\!tensor_k)$ \\
$i\!-\!tensor$	 ::= $ (t\!-\!header\;\; t\!-\!list)$ \\
$t\!-\!list$	 ::= $ NIL\; |\; p\!-\!list\; .\; t\!-\!list$ \\ 
$t\!-\!header$	 ::= $ (t\!-\!name\;\; i\!-\!list)$ \\
$t\!-\!name$	 ::= $ (t_1\;\; t_2\;\; ...\;\; t_k)$ \\ 
$i\!-\!list$	 ::= $ (i_1\;\; i_2\;\; ...\;\; i_l) $
\vskip 1mm
}

\noindent
where $t_1, t_2, ..., t_k$ are basic tensors identifiers, and 
$i_1, i_2, ..., i_l$ are indices (identifiers).
\vskip 2mm

Let us consider an example tensor expression and its representation in
the internal notations.

Let $tt(i,j)$ be a tensor of second order. The internal representation is 
$$
(!\!:\!tensor\; (((tt)\; (i\;\; j))\; (1\; .\; 12))) 
$$
Thus, the tensor expression 
$$
tt(i,j)+tt(j,i) 
$$
will have the internal representation
$$
(!\!:\!tensor\; (((tt)\; (i\;\; j))\; (1\; .\; 21)\; (1\; .\; 12))) 
$$
 
 The most important procedures from this block are described bellow  
in section \ref{UserInt}.
 
 The simplification of tensor expressions is performed by the function
$t\_simp$. The result of this procedure is the canonical form of the 
tensor expression, i.e. the sieved vector $t' \in {\bf L}$.

\subsection{Tensor multiplications}			\label{TMult}
\vskip 0.2cm

The main procedures of the tensor multiplication block are the following:

\begin{itemize}
\item  {$t\_split(tt)$ - splits a term of the tensor expression into the 
       list of basic tensors as factors; }
\item  {$t\_fuse(tf_1,tf_2)$ - combines  tensor 
       factors $tf_1$ and $tf_2$  into  the product.
       This operation is reverse to the previous one;
       } 
\item  {$addmultsym(t_1,t_2)$ - adds symmetry and multiterm linear identity 
        relations generated by the  multiplication to the ${\bf K}$-basis.} 
\end{itemize}

\noindent
Some utilities are available  to work with tensors.

\begin{itemize}
\item {$t\_upright(tt,th)$ - extends the tensor $tt$ to the right with
       respect to the $t$-header $th$;
       } 
\item {$t\_upleft(tt,th)$ - extends the tensor $tt$ to the left with 
       respect to the $t$-header $th$;
       } 
\item {$t\_pri(tt)$ - outputs the tensor $tt$ in the natural form.}
\end{itemize}

\subsection{Dummy indices}

      Dummy relations are created in the process of
evaluation of a tensor expressions. Their number may be very large
and unpredictable in advance. 
Therefore we do not save ${\bf D}$-relations in contrast to 
${\bf S}$ and ${\bf I}$ ones. This leads to the loss of time but saves the
 memory.

 During simplification of tensor expressions we use internal names for
indices. Original names are saved and used in the I/O process. 
Thus, if we have, for example, a dummy index $i$ (really there are two such
names in the expression considered)
then it will be replaced with two internal names: 
$$
\_nn\qquad \mbox{and}\qquad \_mm, 
$$
where $mm=nn+1$. The original name is stored as
a special property of the new ones. 

The dummy block produces relations generated by renamings of dummies. 
The main procedures are the following:

\begin{itemize}
\item  {$adddummy(tt)$ - adds the new relations to the ${\bf K}$-basis; 
       }
\item  {$dl\_get(il)$ - returns the list of dummy indices from the index 
       list $il$;
       }
\item  {$il\_simp(il)$ - replaces original names of the dummy indices
       with their internal names;
       }
\item  {$mk\_dsym(t_1)$ - returns the list of tensor relations with changed  
       dummy indices in each pair;
       } 
\item  {$mk\_ddsym(t_1)$ - returns the list of tensor relations with 
       permuted pairs of dummy indices. 
       }
\end{itemize}

\subsection{Working with ${\bf K}$-bases} \label{WorkWithK}

This block contains the procedures for working with ${\bf K}$-bases.

All ${\bf K}$-bases for various tensor expressions are stored as 
lists in the global variable 
$$
!\!*\!basis ::= (k\!-\!basis_1\; k\!-\!basis_2\; ...)
$$
The structure of the basis is: 
$$
k\!-\!basis ::= (t\!-\!header)\; .\; t\!-\!list 
$$
where the header $t$ and the list $t$ are defined above.

The main procedures are the following:

\begin{itemize}
\item  {$sieve\_pv(pv,b)$ -- sieves the $p$-vector $pv$ using the basis $b$. 
       This procedure is used for the con\-struc\-tion of the ${\bf K}$-basis 
      ("triangle" transformation 
       (\ref{eq:triangK},\ref{eq:Kopt})) and for the simplification of
       tensors expressions (projection to the canonical element
       (\ref{eq:CanProj})). 
       This is the main step of the function $sieve\_t$.;
       }
\item  {$reduce\_pv(pv,qv)$ - reduces the $p$-vector $pv$ with respect 
       to the $p$-vector $qv$. 
       This is the main step of the function $sieve\_pv$.;
       }
\item  {$insert\_pv(pv,b)$ - inserts the $p$-vector $pv$ into the basis $b$.
       This procedure also rearranges $b$ with respect to $pv$. 
       }
\item  {$sieve\_t(tt)$ - sieves the tensor $tt$ using the corresponding  
       ${\bf K}$-basis. The first step of this procedure is generation 
       of relations due to renamings of dummies and the corresponding
       completion of ${\bf K_M}$ up to ${\bf K_D}$-basis.
       }
\end{itemize}

\subsection{Global variables}

In this section we describes the main global variables which allow a user to
control the work. We show the default values in brackets.

\begin{itemize}
\item  {$!\!*\!ppacked(T)$ - are permutations stored in packed form?}
\item  {$!\!*\!debug(NIL)$ - switches the debug output on.}
\end{itemize}

\section{ User's interface} \label{UserInt}

To simplify the user interface, we  restricted the number of 
additional commands. The names of these commands are very similar to the
standard REDUCE ones used in similar cases.

\subsection{$KBASIS$}

The command $KBASIS$ prints the tensor ${\bf K}$-basis. 

The number of vectors in the basis, i.e. the dimension of the
corresponding subspace ${\bf K}$, is typed in the last line of the output.
 Format of this command is: 
$$KBASIS\;\; tt_1,tt_2,...,tt_n;$$
Here $tt_1,tt_2,...,tt_n$ are tensor names. 

To output the ${\bf K}$-basis in the case of the multiplication of two or more 
tensors, it is necessary to use the following format of the command:

$$ KBASIS\;\; t_1(t_2,...,t_k),\ldots ;$$
Here $t_1, t_2, \ldots$ are the names of tensor factors.

If some names have not been declared as tensors the message is produced

\vskip 2mm
\centerline{\it ***** basis1 *** Invalid as tensor: tt}

\subsection{$TENSOR$}

The command $TENSOR$ declares new tensors. 
Format of this command is: 
$$TENSOR\;\; t_1,t_2,...,t_n;$$
Here $t_1,t_2,...,t_n$ are identifiers. The number of indices will be
fixed during the first evaluation of a tensor expressions.

If some names have been declared as tensors already the message is produced:

\vskip 2mm
\centerline{\it +++ tt is already declared as tensor.}

\subsection{$TCLEAR$}				\label{TClear}

The command $TCLEAR$ removes tensors from the list of tensors.
Format of this command is: 
$$TCLEAR\;\; t_1,t_2,...,t_n;$$
Here $t_1,t_2,...,t_n$ are the names of tensor (identifiers).

If some names have not been declared as tensors the message is produced

\vskip 2mm
\centerline{\it +++ xxx is not a tensor.}
\vskip 2mm

{\it Note:} All ${\bf K}$-bases where any of $t_i$ is included as a factor 
will be lost.

\subsection{$TSYM$} \label{tsym}

The command $TSYM$ defines symmetry relations of basic tensors. 
Format of this command is: 
$$TSYM\;\; te_1,te_2,...,te_k;$$
Here $te_1, te_2, ..., te_k$ are linear combinations of basic tensors  
with integer coefficients not containing without dummy indices.

All relations correspond to the left hand side of the symmetry equations. 
For example, for anti\-sym\-metric tensor we have the relation
$AA(i,j)+AA(j,i)=0$. Thus, the corresponding input format is:
$$TSYM\;\; AA(I,J)+AA(J,I);$$

\subsection{Algebraic operations}

The standard algebraic operations are available for tensors:

\begin{tabular}{p{5mm}p{6cm}}
$+$&       sum of tensors;\\
$-$&       difference of tensors or negation;\\
$*$&       multiplication of tensors.
\end{tabular}

We assume that two indices with identical names means the summation over
their values  (the Einstein convention) - they are dummy ones. 
Thus, the multiplication of two tensors may be
either a direct product, or it can contain contractions of dummy indices.

Examples:

\noindent 
$$t_1(i,j)*t_2(j,k);$$

\noindent
$$t_1(i,j)+2*t_2(j,i);$$

\subsection{Switch $DUMMYPRI$}

This switch is controlled by the standard REDUCE commands $ON$/$OFF$. 
It controls the output process for tensor expressions. 
The default value is $OFF$. 

$DUMMYPRI$ prints dummy indices with internal names -- numbers. 
The general rule is:the index $(2k-1)$ is contracted with the index $(2k)$.

Examples\footnote{Here and below REDUCE output is given after the 
arrow "$\Rightarrow$".}:

\vskip 1mm
\noindent
{\small
\begin{tabular}{p{4cm}p{3cm}}
$TENSOR\;\;\; GG;$&  \\
$GG(M,M);$      & $\Rightarrow GG(M,M)$ \\
$ON\;\;\; DUMMYPRI;$&  \\
$GG(M,M);$      & $\Rightarrow GG(M_{41},M_{42})$ 
\end{tabular}
}
\vskip 1mm

\subsection{Switch $SHORTEST$}

This switch is controlled by the standard REDUCE commands $ON$/$OFF$. 
It controls the output process for tensor expressions. 
The default value is $OFF$. 

$SHORTEST$ prints tensor expression in shortest form that was produced
during evaluation. Hoever, the sortest form may be noncanonical.

\vskip 1mm
\noindent
{\small
$TENSOR\;\;\; C;$ \\
$TSYM\;\;\; C(K,L,M)+C(L,M,K)+C(M,K,L);$\\
$C(K,L,M)+C(M,L,K);$ \\ 
$\mbox{\hspace{1cm}} \Rightarrow (-1)*C(L,M,K)+(-1)*C(M,K,L)+C(M,L,K)$ \\
$ON\;\;\; SHORTEST;$ \\
$C(K,L,M)+C(M,L,K);$ \\
$\mbox{\hspace{1cm}} \Rightarrow C(K,L,M)+C(M,L,K)$ 
}
\vskip 1mm

\section{Memory usage}  			\label{UseMem} 

 Let us consider simplification of a tensor expressions with $n$ indices. 
 The rank of the corresponding permutation group is $n$ and the 
 dimension of its group algebra is $n!$. Let us consider two 
 cases: when there are many ${\bf S-I-D}$ relations (so that the dimension
 of the {\bf K} subspace is almost equal to $n!$) and when there are small
 number of ${\bf S-I-D}$ relations (so that the dimension of the {\bf K}
 subspace is small).
 
 In the first case, about $n! \cdot l \cdot k$ Lisp cells are necessary
 to store the full ${\bf K_D}$-basis. 
 Here $l$ is the number of cells needed to store a single term of a $p$-vector,
 and $k$ is the average number of terms in vectors from this ${\bf K_D}$-basis.
The number of terms in these vectors ($k$) is about $2-3$ in practical cases.
The number of terms in the simplified expression is (in practical cases) 
$O(1)$, and can be omitted from this estimate. 
 
In the case of a small set of symmetries and linear identities of basic
 tensors, the number of vectors in the full 
 ${\bf K_D}$-basis is small enough, and can be omitted from this estimate. 
 However, the number of terms in a canonical representative (the expression
 after simplification) will be about $n!$.
 
 Anyway, we have to work with a practically full set of 
 permutations, which contains $n!$ members.
 
 Thus, the minimum computer memory necessary to store the elaborated
 expressions is not less then $n! \cdot l \cdot k$. The typical number of Lisp
 cells necessary to store a single term of a tensor expressions (basic tensor)
 is $4$. The length of each cell is $8$ Byte.  
 The results of calculation for different ranks of the permutation group are
 collected in the table.

\vskip 2mm 
\begin{tabular}{|p{2cm}|p{2cm}|p{2cm}|}
\hline 
Rank of $S_n$	& Number of Mcells	& Memory in Mbyte\\
\hline 
9		& 2.9			& 22.6 \\ 
10		& 29.0			& 226.8 \\ 
11		& 319.3			& 2494.8 \\ 
\hline
\end{tabular} 

\vskip 2mm
2--3 times more memory is necessary in the intermediate steps of calculations.
Modern computers usually equipped with up to $512$ Mbyte memory 
can elaborate tensor expressions with not more then 10 indices with the help 
of the proposed algorithm. However, hardware development is very fast now, and
it will be possible to solve problems with 11 indices with the help 
of our program. Finally, we note that it possible to modify the 
algorithm so that the memory limitations would be not so hard.
However, this advantage is compensated by a significant increase of the
execution time. To summarize this section, we conclude that the absolute limit
for the group algebra approach developed in this work is 12 indices.

\section{Examples}  

\subsection{Symmetric and antisymmetric tensors}

At the beginning, let us declare the names of basic tensors:

\vskip 0.2cm
\noindent
$tensor\;\; s2,a3,v1,v2,v3;$
\vskip 0.2cm

By using $TSYM$ command we introduce the {\bf S-I} relations of the basic
tensors:

\vskip 0.2cm
\noindent
{\small
\begin{tabular}{lll}
$tsym $&$ s2(i,j)-s2(j,i),$ & \% Symmetric  \\
$     $&$ a3(i,j,k)+a3(j,i,k),$ & \% Antisymm. \\
$     $&$ a3(i,j,k)-a3(j,k,i);$ & \\
\end{tabular}
}
\vskip 0.2cm

Let us output the {$\bf K_0$}-bases constructed for the tensors $a_2$ and $s_2$

\vskip 0.2cm
{\small
\noindent
$kbasis\;\; s2,a3;$
\vskip 0.2cm

\begin{tabular}{ll}
$\Rightarrow$ & $s2(j,i) + (-1)*s2(i,j)$\\
$\Rightarrow$ & $1$\\
              & \\
$\Rightarrow$ & $a3(k,i,j) + a3(j,i,k)$\\
$\Rightarrow$ & $a3(k,j,i) + (-1)*a3(j,i,k)$\\
$\Rightarrow$ & $a3(i,k,j) + (-1)*a3(j,i,k)$\\
$\Rightarrow$ & $a3(i,j,k) + a3(j,i,k)$\\
$\Rightarrow$ & $a3(j,k,i) + a3(j,i,k)$\\
$\Rightarrow$ & $5$
\end{tabular}
}
\vskip 0.2cm

Now we are ready to simplify tensor expressions. Some examples are:

\vskip 2mm
{\small
\noindent
\begin{tabular}{ll} 
$s2(i,j) + s2(j,i);$ & $\Rightarrow  2*s2(i,j)$ \\
$a3(i,j,k)*s2(i,j);$ & $\Rightarrow  0$ \\
\end{tabular}

\noindent
\begin{tabular}{ll} 
$a3(i,j,k)*v1(i)*v2(j)*v1(k);$ & $\Rightarrow  0$ 
\end{tabular}
}
\vskip 2mm

   Sometimes one can get a 'strange' output if one will not be careful with
the input. For example,

\vskip 2mm
\noindent
\begin{tabular}{ll} 
$x:=s2(i,i);$ & $\Rightarrow  x:=s2(i,i)$\\
$x*v1(i);$ & $\Rightarrow  s2(i,i)*v1(i)$\\
\end{tabular}
\vskip 2mm

\noindent
{}From the standard point of view, the second output is incorrect due to
the fact that three indices with the same name are present. However, the input
has not been recognized as an error. If one switches on the flag

\vskip 2mm
\noindent
$on\;\;\; dummypri;$
\vskip 2mm

\noindent
and then repeats the previous input then one gets the following output:

\vskip 2mm
\noindent
\begin{tabular}{ll}
$x*v1(i);$	& $\Rightarrow  s2(i_{23},i_{24})*v1(i)$
\end{tabular}
\vskip 2mm

\noindent
Hence, the first two $i$'s are dummies and the last one is a free index.

  If a user would like to output the ${\bf K}$-basis of the product 
of the tensors $s_2$ and $a_3$, the following format of the 
command $KBASIS$ has to be used:

\vskip 2mm
{\small
\noindent
$kbasis\;\; s2(a3);$

\vskip 2mm 
\noindent
\begin{tabular}{lll}
&	$\Rightarrow$ & $a3(j,i,k)*s2(i,j) + a3(k,i,j)*s2(j,i)$ \\ 
&	$\Rightarrow$ & $a3(j,i,k)*s2(j,i) + a3(k,i,j)*s2(j,i)$ \\ 
&	.....& \\ 
&	$\Rightarrow$ & $110$
\end{tabular}
}

\subsection{Riemann tensor}

Let us introduce the Riemann tensor and the standard set of its {\bf S-I}
relations:

$tensor\;\; ri;$ \\ 
$tsym\;\; ri(i,j,k,l) + ri(j,i,k,l);$ \\ 
$tsym\;\; ri(i,j,k,l) + ri(i,j,l,k);$ \\ 
$tsym\;\; ri(i,j,k,l) + ri(i,k,l,j) + ri(i,l,j,k);$ 
\vskip 2mm

The ${\bf K_0}$-basis consists of 22 vectors (see TEST RUN OUTPUT) and the 
full vector space has $4!=24$ dimensions. Thus, any expressions which are
linear combinations of Riemann tensors with permuted indices can be 
simplified to expressions containing only two basic
tensors\footnote{This simplification has no relation to the
number of independent components of the Riemann curvature tensor in
space-time of various dimensions}. 

This set of properties leads us to the very important symmetry property 
of Riemann tensor with respect to the exchange of pairs of indices:

$ri(i,j,k,l) - ri(k,l,i,j); \qquad \Rightarrow 0$
\vskip 2mm

Let us consider some more examples,

\vskip 2mm
\noindent
$ri(m,n,m,n)-ri(m,n,n,m)$ \\
$\mbox{\hspace{4cm}} \Rightarrow 2*ri(m,n,m,n)$. 
\vskip 2mm

Any tensors expressions consists of Rieman tensors may be expressed
through summ of 2 ones:
 
\vskip 2mm
\noindent
$ri(i,j,k,l)+ri(j,k,l,i)+ri(k,l,i,j)+ri(l,i,j,k);$ \\
$\mbox{\hspace{1cm}} \Rightarrow (-2)*ri(l,j,i,k)+4*ri(l,i,j,k)$
\vskip 2mm

A more complicated example with multiplication of the Riemann tensor and
the antisymmetric tensor $a2$ is given in the section {\bf TEST RUN OUTPUT}.

\section*{Acknowledgements}

  The authors are grateful to A.Grozin for useful discussions.

  This work was supported by Russian Foundation for Fundamental Research
  (grant 93-02-14428).

\onecolumn
\vskip 1cm
{\centerline {\Large \bf TEST RUN OUTPUT} }

\begin{verbatim} 
%*********************************************************************
%                          ATENSOR  TEST  RUN.
%
%                       V.A.Ilyin & A.P.Kryukov
%                  E-mail:   ilyin@theory.npi.msu.su
%                          kryukov@theory.npi.msu.su
%
%                Nucl. Phys. Inst., Moscow State Univ. 
%                        119899 Moscow, RUSSIA
%*********************************************************************

% First of all we have to load the ATENSOR program using the one of the
% following command:
%       1) in "atensor.red"$         % If we load source code
%       2) load atensor$             % If we load binary (compiled) code.
load atensor;

(atensor)

% To control of total execution time clear timer:
showtime;

Time: 0 ms

% Switch on the switch TIME to control of executing time 
% for each statement.
%on time$

% Let us introduce the antisymmetric tensor of the second order.
tensor a2;

% The antisymmetric property can be expressed as:
tsym a2(i,j)+a2(j,i);

% The K-basis that span K subspace is:
kbasis a2;

a2(i,j) + a2(j,i)
1

% Let us input very simple example:
a2(k,k);

0

% By the way the next two expressions looks like different ones:
a2(i,j);

a2(i,j)

a2(j,i);

a2(j,i)

% But the difference of them has a correct value:
a2(j,i)-a2(i,j);

2*a2(j,i)

% Next examples. For this purpose we introduce 3 abstract 
% vectors - v1,v2,v3:
tensor v1,v2,v3;

% The following expression equal zero:
a2(i,j)*v1(i)*v1(j);

0

% It is interest that the result is consequence of the equivalence 
% of the name of tensors.

% While the next one - not:
a2(i,j)*v1(i)*v2(j);

a2(i,j)*v1(i)*v2(j)

% Well. Let us introduce the symmetric tensor of the second order.
tensor s2;

tsym s2(i,j)-s2(j,i);


% Their K-basis look like for a2 excepted sign:
kbasis s2;

s2(j,i) + (-1)*s2(i,j)
1

% Of course the contraction symmetric  and antisymmetric tensors 
% equal zero:
a2(i,j)*s2(i,j);

0

% By the way, the next example not so trivial for computer...
a2(i,j)*a2(j,k)*a2(k,i);

0

% Much more interesting examples we can demonstrate with the
% the tensor higher order. For example full antisymmetric tensor
% of the third order:
tensor a3;

% The antisymmetric property we can introduce through the 
% permutation of the two first indices: 
tsym a3(i,j,k)+a3(j,i,k);

% And the cyclic permutation all of them:
tsym a3(i,j,k)-a3(j,k,i);

% The K basis of a3 consist of 5 vectors:
kbasis a3;

a3(k,i,j) + a3(j,i,k)
a3(k,j,i) + (-1)*a3(j,i,k)
a3(i,k,j) + (-1)*a3(j,i,k)
a3(i,j,k) + a3(j,i,k)
a3(j,k,i) + a3(j,i,k)
5

% In the beginning some very simple examples:
a3(i,k,i);

0

a3(i,j,k)*s2(i,j);

0

% The full symmetric tensor of the third order may be introduce
% by the similar way: 
tensor s3;

tsym s3(i,j,k)-s3(j,i,k);

tsym s3(i,j,k)-s3(j,k,i);

kbasis s3;

s3(k,j,i) + (-1)*s3(i,j,k)
s3(k,i,j) + (-1)*s3(i,j,k)
s3(j,k,i) + (-1)*s3(i,j,k)
s3(j,i,k) + (-1)*s3(i,j,k)
s3(i,k,j) + (-1)*s3(i,j,k)
5

% The next examples demonstrate some calculation with them:
s3(i,j,k)-s3(i,k,j);

0

s3(i,j,k)*a2(i,j);

0

a3(i,j,k)*s2(i,j);

0

s3(i,j,k)*a3(i,j,k);

0

% Now we consider very important physical case - Rieman tensor:
tensor ri;

% It has the antisymmetric property with respect to the permutation
% of the first two indices:
tsym ri(i,j,k,l) + ri(j,i,k,l);

% It has the antisymmetric property with respect to the permutation
% of the second two indices:
tsym ri(i,j,k,l) + ri(i,j,l,k);

% And the triple term identity with cyclic permutation the 
% third of them:
tsym ri(i,j,k,l) + ri(i,k,l,j) + ri(i,l,j,k);

% The corresponding K basis consist of 22(!) vectors:
kbasis ri;

ri(l,k,i,j) + (-1)*ri(j,i,k,l)
ri(l,k,j,i) + ri(j,i,k,l)
ri(l,i,k,j) + (-1)*ri(j,k,i,l)
ri(l,i,j,k) + ri(j,k,i,l)
ri(l,j,k,i) + (-1)*ri(j,k,i,l) + ri(j,i,k,l)
ri(l,j,i,k) + ri(j,k,i,l) + (-1)*ri(j,i,k,l)
ri(k,l,i,j) + ri(j,i,k,l)
ri(k,l,j,i) + (-1)*ri(j,i,k,l)
ri(k,i,l,j) + (-1)*ri(j,k,i,l) + ri(j,i,k,l)
ri(k,i,j,l) + ri(j,k,i,l) + (-1)*ri(j,i,k,l)
ri(k,j,l,i) + (-1)*ri(j,k,i,l)
ri(k,j,i,l) + ri(j,k,i,l)
ri(i,l,k,j) + ri(j,k,i,l)
ri(i,l,j,k) + (-1)*ri(j,k,i,l)
ri(i,k,l,j) + ri(j,k,i,l) + (-1)*ri(j,i,k,l)
ri(i,k,j,l) + (-1)*ri(j,k,i,l) + ri(j,i,k,l)
ri(i,j,l,k) + (-1)*ri(j,i,k,l)
ri(i,j,k,l) + ri(j,i,k,l)
ri(j,l,k,i) + ri(j,k,i,l) + (-1)*ri(j,i,k,l)
ri(j,l,i,k) + (-1)*ri(j,k,i,l) + ri(j,i,k,l)
ri(j,k,l,i) + ri(j,k,i,l)
ri(j,i,l,k) + ri(j,i,k,l)
22

% So we get the answer for any expressions with 3 and more terms of 
% Rieman tensors with not more then 2 terms. For example:
ri(i,j,k,l)+ri(j,k,l,i)+ri(k,l,i,j)+ri(l,i,j,k);

(-2)*ri(l,j,i,k) + 4*ri(l,i,j,k)
 
% This three identities leads us to very important symmetry property with
% respect to exchange of pairs indices:
ri(i,j,k,l)-ri(k,l,i,j);

0

% Let us start with simple example:
ri(m,n,m,n)-ri(m,n,n,m);

2*ri(m,n,m,n)

% Much more complicated example is:
a2(m,n)*ri(m,n,c,d) + a2(k,l)*ri(c,d,l,k);

0

% The answer is trivial but not so simple to obtain one.

% The dimension of the full space is 6! = 720.
% The K basis consists of 690 vectors (to reduce output we 
% commented the last statement):
%kbasis ri(a2);

% One else nontrivial examples with Riemann tensors:
(ri(i,j,k,l)-ri(i,k,j,l))*a2(i,j);

 a2(i,j)*ri(i,j,k,l)
---------------------
          2

%***************** END OF TEST RUN ************************
% The total execution time is:
showtime;

Time: 196940 ms  plus GC time: 10670 ms

$

END$

\end{verbatim} 

\end{document}